\documentclass[11pt,a4paper]{article}
\pdfoutput=1

\usepackage{jinstpub} 
                     
\usepackage[table,xcdraw]{xcolor}

\usepackage{amssymb}
\usepackage{amsmath}
\usepackage{subfig}

\title{\boldmath Neutrinos Angra experiment: commissioning and first operational measurements}

\author[a,1]{H. P. Lima Jr \note{Corresponding author.},}
\author[d]{J. A. M. Alfonzo,}
\author[a]{J. C. Anjos,}
\author[a]{G. Cernicchiaro,}
\author[c]{P. Chimenti,}
\author[b]{I. A. Costa,}
\author[b]{M. P. Dias,}
\author[d]{P. C. M. A. Farias,}
\author[b]{A. Fernandes Junior,}
\author[e]{G. P. Guedes,}
\author[f]{L. F. G. Gonzalez,}
\author[f]{E. Kemp,}
\author[b]{G. S. Lopes,}
\author[d]{J. Marcelo,}
\author[b]{M. L. Migliorini,}
\author[b]{R. A. N\'obrega,}
\author[d]{I. M. Pepe,}
\author[d]{D. B. S. Ribeiro,}
\author[b]{D. M. Souza,}
\author[d]{L. R. Teixeira,}

\affiliation[a]{Centro Brasileiro de Pesquisas F\'{i}sicas, Rio de Janeiro, RJ, Brazil}
\affiliation[b]{Universidade Federal de Juiz de Fora, Juiz de Fora, MG, Brazil}
\affiliation[c]{Universidade Estadual de Londrina, Londrina, SC, Brazil}
\affiliation[d]{Universidade Federal da Bahia, Salvador, BA, Brazil}
\affiliation[e]{Universidade Estadual de Feira de Santana, Feira de Santana, BA, Brazil}
\affiliation[f]{Universidade Estadual de Campinas, Campinas, SP, Brazil}

\emailAdd{hlima@cbpf.br}

\abstract{The Neutrinos Angra Experiment has completed a major step by finishing the commissioning of the detector and the data acquisition system at the experimental site located in the Angra dos Reis nuclear power plant. The experiment, consisting of a water-based detector and associated electronics, was designed with the goal of detecting the electron antineutrinos produced by the nuclear reactor. The detection is possible due to the Inverse Beta Decay, where the final products in the water are photons in the UV-to-visible range of the spectrum. The assembled detector comprises three active volumes filled with water: (i) a cubic detector (Target) for electron antineutrinos, covered by 32 8-inch photomultiplier tubes (PMTs), (ii) a lateral layer surrounding the Target (Lateral veto) equipped with 4 PMTs and (iii) a third volume covering the top of both (Top veto), also equipped with 4~PMTs. In the present document the main features of the detector assembly as well as the integration of the readout electronics on-site are reported. Finally, some operational characteristics are shown based on analysis of the first measurements performed with the fully working detector.}


\begin{document}
\maketitle
\flushbottom

\section{Introduction}
Nuclear reactors have been crucial to experimental neutrino physics as they are copious man-made sources of neutrinos. In addition to having made possible the confirmation of the neutrino hypothesis in 1956 \cite{cowan}, other experiments have been built since then inside such facilities. By the end of 90s and early 2000s, reactor experiments like Palo Verde \cite{palo} and CHOOZ \cite{chooz} put upper limits on the value of $\theta_{13}$. Soon thereafter, the KamLAND experiment \cite{kamland} pinned down the large mixing angle (LMA) solution to the solar neutrino problem again using the antineutrino flux from distant reactors. Almost a decade later, Daya Bay \cite{dbay}, Double Chooz \cite{dchooz} and RENO \cite{reno} experiments considerably increased the knowledge on the neutrino oscillation phenomenon by the precise measurement of the mixing angle $\theta_{13}$.

The neutrino emission from nuclear reactors is closely related to the fission of heavy nuclei taking place inside the reactor: each fission contributes with a well-known fraction of the total released energy and is followed by the emission of neutrinos.  By using a neutrino detector to monitor the flux, one can estimate the fission's rate and thus the energy released by the reactor, i.e., the neutrino flux is directly proportional to the power of the reactor.

The proposal to use neutrinos for remote monitoring of the thermal power of nuclear reactors was first considered in the mid-1970s \cite{Mik,BoroMik}. One of the first demonstration experiments was performed in a neutrino laboratory located in the nuclear power plant at Rovno - Ukraine. The relationship between neutrino counting rate and the reactor activity was clearly shown, so the neutrino radiation could be, in principle, used for such purposes \cite{Rovno1,Rovno2}.

This scenario described above opens up solid perspectives for the use of neutrinos as reliable probes of the physical processes in which they participate. Thus, a neutrino detector can monitor parameters related to the activity of nuclear reactors that are crucial for checking items of the non-proliferation safeguards dictated by the International Atomic Energy Agency (IAEA).

An important point in the use of neutrino detectors for nuclear reactor monitoring, especially regarding the verification of safeguards, is the possibility to remotely check the reactor activity, avoiding the need for operation and intrusion in the containment area or other restricted access areas in the nuclear power plant. Summaries of the worldwide effort on the use of antineutrinos detectors for safeguards purposes can be found in \cite{safeguards} \cite{korea} and references therein. 

A neutrino experiment at the Brazilian nuclear power plant in Angra dos Reis was initially considered for studies of neutrino oscillations \cite{angra-1st}. After the alignment of the groups around the three proposals that would be implemented (Double Chooz, RENO and Daya Bay), the Brazilian group decided that there was a great opportunity to perform an experiment in the premises of the plant, even if of smaller dimensions and with another purpose, in this case the verification of non-proliferation safeguards. This is the origin of the Neutrinos Angra ($\nu$-Angra) experiment, discussed in this paper.

The long term goal of the $\nu$-Angra experiment is to develop a reliable and cost-effective technology to routinely monitor the nuclear reactor power and possibly the neutrino spectral evolution. The stability of the data acquisition is a fundamental step in this direction. This goal is well aligned with the safeguards and non-proliferation demands of the IAEA. In the current commissioning phase, we have fully validated the electronics, which is operating within the desired stability to run for years of data taking. In this paper we report the main features of the detector assembly as well as the integration of the readout electronics on-site and results from the analysis of initial data taking.

A common challenge for all neutrino experiments, using reactors or not, is the background radiation -  mainly neutrons and cosmic muons - that can mimic signals with similar features of those expected from neutrino detection. To suppress the background level, the usual solution is to install the detectors in large underground caverns \cite{chooz,dchooz} using the rock and soil overburden as a natural shield. Several measurements have shown that the vertical muon intensity can be reduced $10^{4}$ times for $10^{2}~hg~cm^{-2}$ depth of rock \cite{grieder}. Achieving a Signal-To-Noise Ratio (signals produced by neutrinos against neutrino-like signals caused by other particles) high enough in order to make possible to monitor the burning process of nuclear reactors is the main  challenge (if not the biggest) for the $\nu$-Angra experiment, since the detector is assembled on the surface, one of the points of the agreement with the plant operator to conduct the research inside the Angra dos Reis nuclear complex. 
In order to handle with the background, the main tools that will be used are: a cosmic rays VETO system surrounding the outermost part of the detector target volume, and discrimination techniques in data analysis based on the measurements of time and energy of events.

The purpose of this paper is to describe the commissioning of the experiment, which started in September 2017 and finished in the second half of 2018. The paper is organized as follows: section 2 provides an overview of the whole detector - including the neutrino TARGET and the two VETO volumes - and the Readout Electronics (RE) developed to readout and storage of the signals; section 3 presents the detector assembling procedure in the nuclear reactor site; section 4 covers a brief description of the RE, its installation in the site and the initial tests; the first measurements, and respective data analysis, with the full detector and readout electronics in operation are presented in section 5. Finally, conclusions and perspectives close the paper in section 6.

\section{The $\nu$-Angra Detector and the Readout Electronics}
\subsection{Detector Overview}
\begin{figure}[ht!]
	\centering
	\includegraphics[scale=0.3]{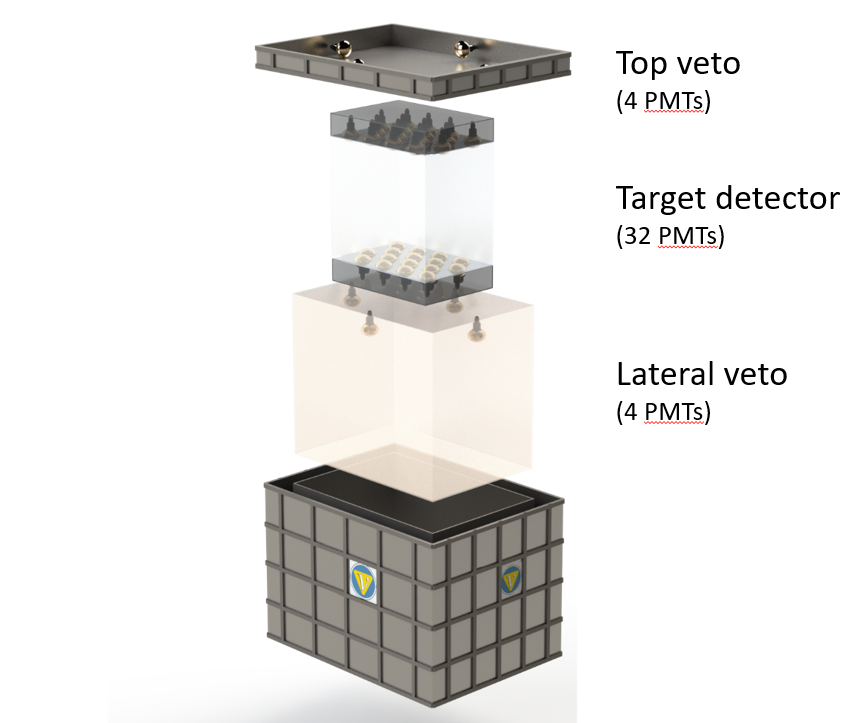}\\
	\caption{Exploded view of the $\nu$-Angra detector.}
	\label{fig:detector}
\end{figure}

The $\nu$-Angra experiment intends to develop a system, based on a water Cherenkov detector, capable of measuring the antineutrino flux emitted by reactors. The Angra dos Reis nuclear site (which houses a 3764~MW power reactor named Angra II) is being used to assess the proposed technology. With the reactor in steady-state operation, the neutrino flux produced by that unit is estimated as $1.21\times10^{20}~s^{-1}$ \cite{connie}. The electron antineutrino flux can be used to perform a non-invasive monitoring of the reactor activity as well as to estimate the thermal power produced at the reactor core. For the 1~ton Target detector, the expected rate is around 5000 events per day for a distance of 25~m from the reactor core, enabling the experiment to investigate the potential of antineutrino detection for safeguards applications. The antineutrino interactions are from inverse $\beta$ decays taking place inside the Target detector, a cubic volume filled with 1340.28 l of $GdCl_{3}$ doped water (0.2~\%). The water volume is contained by a 0.90~m$~\times~$1.46~m$~\times~$1.02~m (height-length-width) plastic tank surrounded by 32 eight-inch photomultiplier tubes (PMTs), type Hamamatsu R5912 \cite{pmt}, with 16 installed on the bottom and 16 on the top of the volume, as is shown in figure \ref{fig:detector}. The six internal faces of the target volume are covered by sheets of a diffuse reflector membrane (0.5~mm DRP), which provides reflectance greater than 99.0~\% for 400~nm wavelength light.

The vessel on the top of figure \ref{fig:detector} is the Top VETO detector, designed to detect and generate a trigger pulse whenever a cosmic muon crosses the volume. Four PMTs, of the same model as the one in the Target detector, are fixed on the middle of each side, pointing to the center of its volume. The Top VETO tank is filled with pure water, has dimensions 0.28~m$~\times~$2.66~m$~\times~$2.02~m (h-l-w) and has all the internal faces covered with the Tyvek\footnote{Trademark of DuPont.} reflector material (providing reflectance greater than 97.0~\%). Surrounding the Target detector, there is a first water layer with 12.00~cm thickness. In this part of the detector - called here the Lateral VETO - there are four PMTs fixed in the middle of each side on the top with the photocatode pointing to the bottom. The Lateral VETO corresponds to the third part - from top to bottom - in figure \ref{fig:detector}. The Top and the Lateral VETO layers (each one covered by four photomultipliers), and a dedicated circuitry, form the VETO system, which is responsible for detecting cosmic ray particles that might hit the Target detector volume. Whenever two or more PMTs fire at the same time, a veto window is generated blocking any trigger signal that could eventually be generated due to hits in the Target detector. 

Finally, the outermost part of the detector, which surrounds the VETO Lateral, is 14.50 cm thick on two opposite sides and 22.5 cm on the other two sides. It is completely filled with pure water and used as shield against background neutrons.         
 
\subsection{Readout Electronics}
\begin{figure}[ht!]
	\centering
	\includegraphics[scale=1.0]{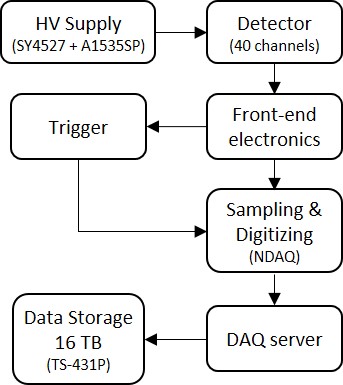}\\
	\caption{Overview of the Readout Electronics installed on the nuclear power plant.}
	\label{fig:daq_overview}
\end{figure}

In addition to the detector itself, a complete data acquisition system has been designed and integrated for the experiment in order to perform tasks, such as: biasing of the PMTs (High Voltage power supply), amplification of PMT output signals (Front-End electronics), sampling and digitization of the signals (NDAQ electronics), online events selection (Trigger system) and local data storage (commercial \emph{Network Attached Storage}). An overview of the Readout Electronics with those systems is the diagram shown in figure \ref{fig:daq_overview}.

The high voltage system and the local computing infrastructure - DAQ (Data Acquisition System) servers and local data storage - are commercial off-the-shelf parts in the readout electronics. The High Voltage Power Supply used to bias the 40 PMTs is a commercial mainframe-based system from CAEN. It is composed of one SY4527 mainframe housing two units of the A1535SP 24-channel positive voltage module with SHV-type output connectors. The HV system is remotely controlled through an Ethernet connection and a simple terminal-style user interface. The local data storage is a \emph{network attached storage} unit, model TS-431P (QNAP), assembled with four 4~TB disks. Data read from the detector is continuously written to those disks and subsequently transferred to two larger and permanent data servers: the primary one installed at CBPF, in Rio de Janeiro, and a mirror server installed at Unicamp, in Campinas.

\subsubsection{Front-end electronics}
\begin{figure}[ht!]
	\centering
	\includegraphics[scale=0.33]{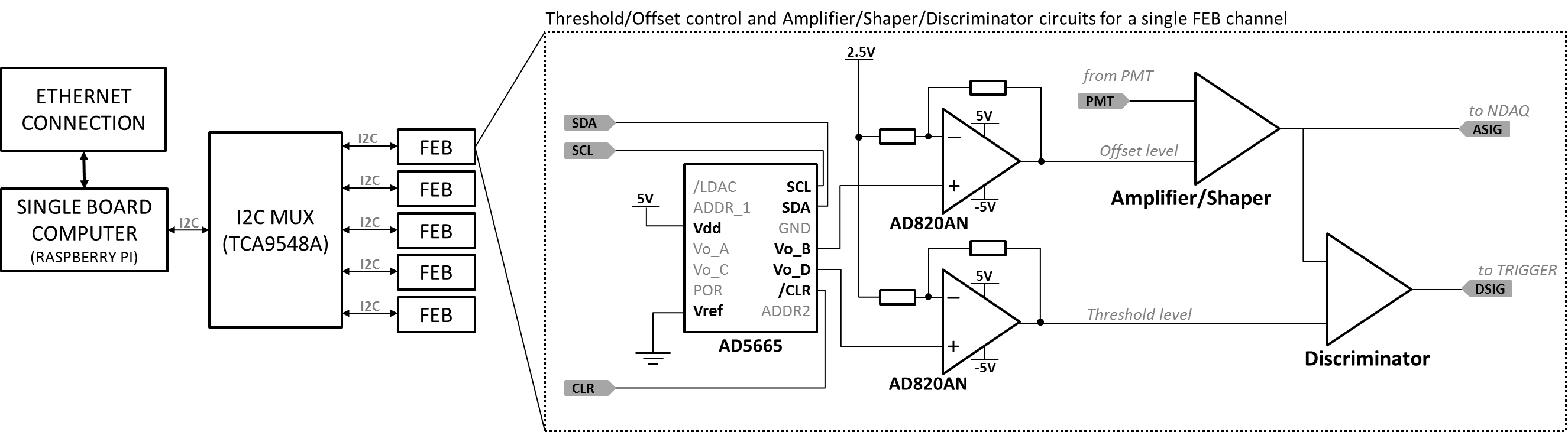}\\
	\caption{Front-end electronics channel.}
	\label{fig:fee_schematics}
\end{figure}

In order to condition the PMTs output signals to the digitizing electronics (NDAQ modules), and to inform which channels have been fired to the trigger system, a custom front-end circuitry has been developed \cite{fee_1}. The front-end sensitivity was measured to be 71.5 $\pm$ 0.9 mV (output signal peak value) per photoelectron and it starts saturating after achieving an output peak amplitude of 1.4 Volts; therefore, each channel can process up to about 20 photoelectrons without linearity loss. Five front-end boards (FEB) are used in the experiment, each one with eight independent channels. A channel is composed of a four-stage amplification/shaper circuit, a discriminator circuit and a control system. The former prepares the signal to the digitizer, the output of the second is delivered to the trigger system and the latter allows remote adjustment of the offset of the analog signal and the threshold of the discriminator circuit according to the experiment requirements. Therefore, as shown in figure \ref{fig:fee_schematics}, each channel offers two output signals: an analog (ASIG) and a discriminated (DSIG) signal.

The control system, also shown in figure \ref{fig:fee_schematics}, is based on an $I^2C$ controllable DAC chip (AD5665), placed on the FEB, and a single-board computer (RaspberryPI) connected to the network to allow accessing all the DAC (Digital-to-Analog Converter) chips remotely. To access all the FEBs, an $I^2C$ multiplexer (TCA9548A) is used. Each DAC provides four output voltages (only two of them are shown in figure \ref{fig:fee_schematics} for simplification) and, therefore, to control all the eight offset and eight threshold levels of a FEB, four DAC chips are used per board. All board functionalities and channels were tested and calibrated in the laboratory before installing the boards in the experiment container. Once they were installed in the nuclear site, all channels had their thresholds lowered until reaching a rate of at most 3000 pulses per second at the discriminator output. The average of the resulting threshold values was estimated to be approximately 70 mV, hence within the single-photoelectron region.

\subsubsection{Digitizer}
The detector signals, amplified and shaped by the front-end electronics, are sent to a VME-based system able to sample and digitize the signals whenever it receives a trigger pulse. The system is composed of a commercial \emph{single board computer} module, model MVME3100 (Artesyn Embedded Technologies), a commercial \emph{Fan-In Fan-Out} module, model V976 (CAEN), and five NDAQ modules, specifically designed for the experiment \cite{ndaq}. The single board computer works as the readout processor of the digitizer cards. It hosts and executes a C language program that control and read, during data taking, the NDAQ cards installed on the mainframe. The fan-in fan-out module is used to distribute the digital trigger pulse coming from the Trigger system to the five NDAQ modules.

\begin{figure}[ht!]
	\centering
	\includegraphics[scale=0.4]{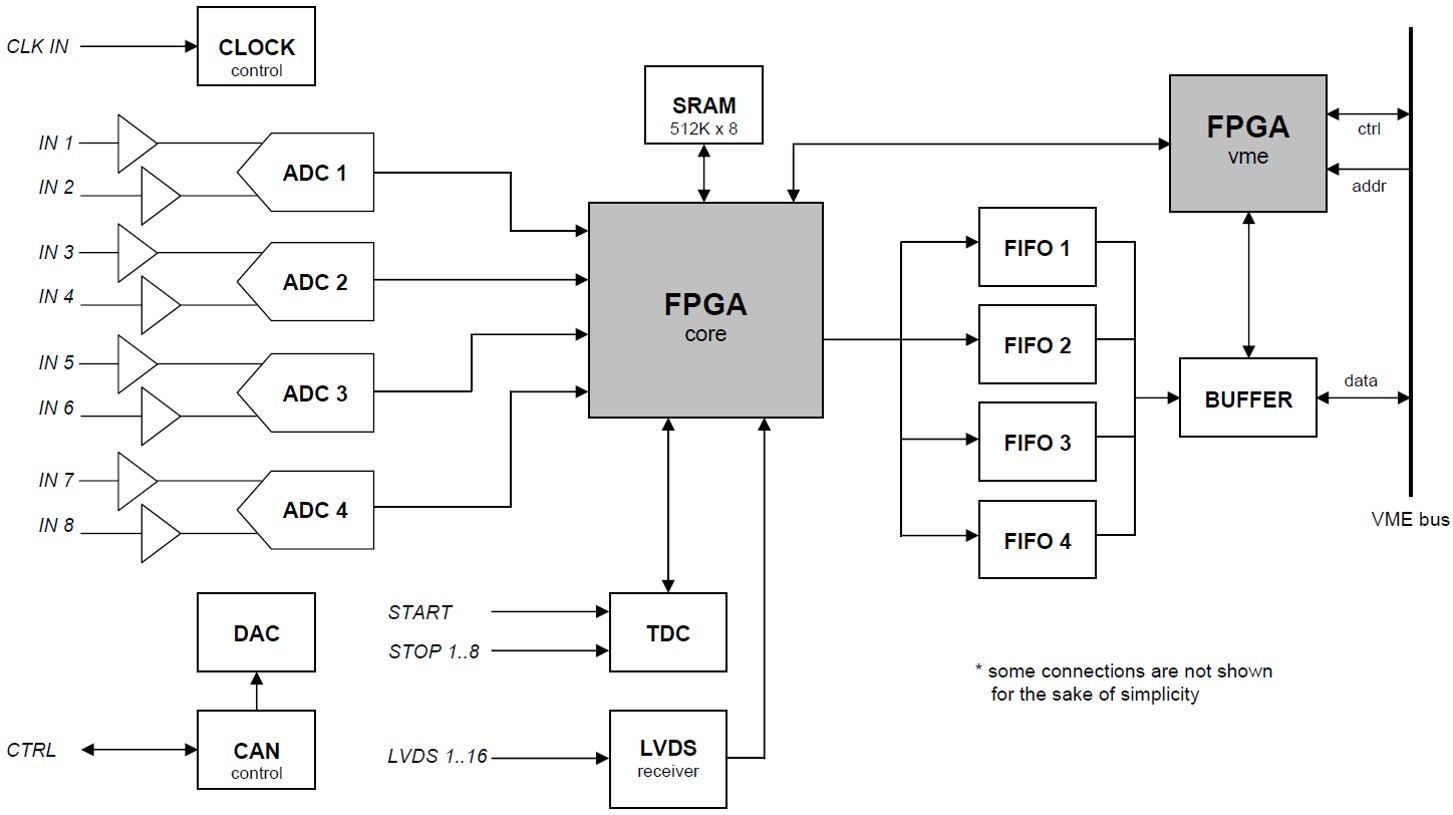}\\
	\caption{NDAQ data acquisition module design.}
	\label{fig:ndaq_diag}
\end{figure}

Each NDAQ module has eight ADC (Analog-to-Digital Converter) channels working at 125~MHz sampling rate with 10~bits vertical resolution, as shown in figure \ref{fig:ndaq_diag}. The ADC output samples are sent to an FPGA (Field Programmable Gate Array), where a circuit based on two FIFO memories connected in series is programmed. This approach is used in order to store pulse samples before and after the rising edge of the trigger pulse, otherwise only samples after the trigger instant would be read. It is worth noting that typical pulses coming out of the photomultiplier tubes feature spectral content up to 160~MHz, at the point where the spectrum falls -20~dB. After the front-end electronics, the signal bandwidth falls to less than 50~MHz and therefore can be properly converted at 125 Mega samples per second. 

\subsubsection{Trigger}
\begin{figure}[ht!]
	\centering
	\includegraphics[scale=0.8]{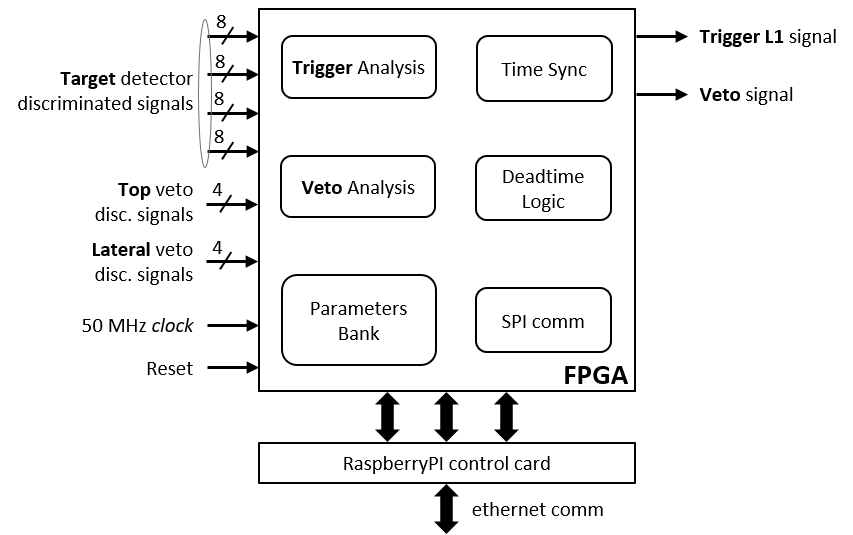}\\
	\caption{Trigger electronics overview.}
	\label{fig:trigger_diag}
\end{figure}

In the occurrence of an event, the front-end output signals are digitized by the NDAQ modules and sent to on-board FIFO memories waiting for a trigger decision. If the event is selected by the Trigger System \cite{trigger}, shown in figure \ref{fig:trigger_diag}, the corresponding data is transferred to the experiment data storage unit for future analysis. For a fast trigger decision, the selection algorithm was developed to be implemented in a dedicated FPGA. For remote configuration and upgrade of its firmware, a RaspberryPI card with Ethernet connection has been integrated to the FPGA circuit.

The Trigger board receives as input the 40 discriminated signals from the discriminator section of the Front-end modules: 32 signals from the Target detector PMTs, 4 from the Lateral Veto and 4 from the Top Veto. Two output connectors deliver the results of the trigger logic processing: the overall Trigger L1 signal, a high level pulse that indicates whenever a minimum of PMT units have been fired at the same time window, and the Veto signal, which informs that a veto condition was found. The Veto condition, as the name suggests, blocks any Trigger L1 pulse during a short period (currently set to 2.5~$\mu$s) after the rising edge of the Veto signal.

In the current design, the Trigger L1 condition is based only on multiplicity, which means that the Trigger L1 pulse is active whenever a minimum number of PMTs are fired in the Target detector inside a predefined time window. The same concept applies to the Veto system. 

A number of configuration registers that define the Trigger functionalities are available and can be remotely configured. The values of those registers are defined in a text file stored in the RaspberryPI control card. In order to perform any change on the registers, one only needs to edit that file. A list of the most relevant registers are given below::
\begin{itemize}
	\setlength\itemsep{0.1em}
	\item minimum number of PMTs fired in the Target detector to generate a Trigger L1 pulse;
	\item minimum number of PMTs fired in the Lateral veto to generate a Veto condition;
	\item minimum number of PMTs fired in the Top veto to generate a Veto condition;
	\item time window length to accept Target PMTs discriminated pulses;
	\item time window length for blocking Trigger L1 pulses in a Veto condition;
	\item blocking mask for each PMT channel in the whole detector (40 bits).
\end{itemize}

\section{Assembling at the Experimental Site}

\subsection{Detector Assembling}

The final assembly of the detector was planned to be done just outside of its final position, inside the container of the project, located 2~m away from the wall of the nuclear reactor building. After an analysis of the best course of action to be taken, it was decided to prepare the Top veto first, due to its relative simplicity when compared to the Target detector tank; then the Target, and finally the Lateral veto were mounted in that order because the latter needed the target inside it to be instrumented. All volumes were already internally coated with Tyvek prior to the transportation to the nuclear reactor site in order to facilitate the assembly.

\begin{figure}[ht!]
	\centering
	\includegraphics[scale=1.0]{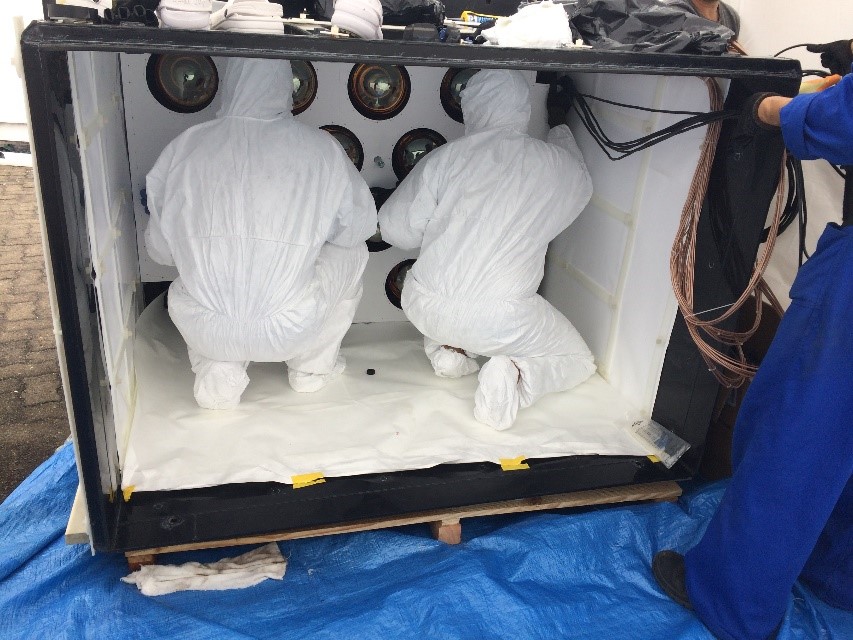}\\
	\caption{Bottom of the Target detector being instrumented in the nuclear power plant.}
	\label{fig:target_assembly}
\end{figure}

\begin{figure}[ht!]
	\centering
	\includegraphics[scale=1.0]{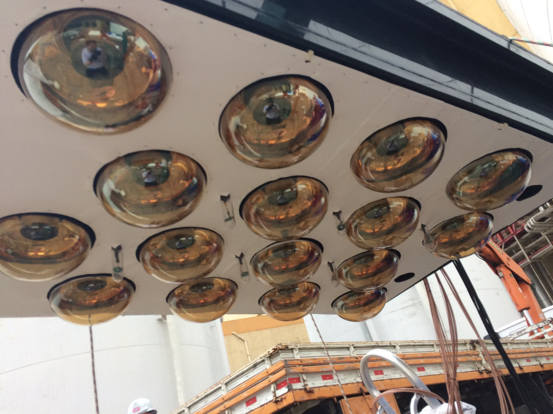}\\
	\caption{Target detector lid with the 16 PMTs and 6 LEDs installed.}
	\label{fig:lid}
\end{figure}

The Top veto was instrumented with 4 PMTs and a calibration LED. To equip the Target detector, it was laid sideways so that two researchers could enter and fit in the 16 bottom PMTs along with 6 calibration LEDs, as shown in figure \ref{fig:target_assembly}. Finally the Target detector tank was carefully brought to its nominal position and externally coated with Tyvek. Its lid was lifted and the 16 top PMTs and 6 LEDs were placed. Figure \ref{fig:lid} shows the Target detector lid fully instrumented just before lowering it down to its final position.

With the help of the nuclear power plant personnel, the Target, now complete, was lifted and placed inside the Lateral veto volume. The 4 PMTs of the Lateral veto were mounted on its main cover plates, upside down. The main plates were covered with secondary plastic plates and a canvas to block light leakage.

The Top veto was lifted and placed upon the Lateral veto tank. To finish the detector assembling, another canvas was placed on top of the detector and it was moved inside the container. 

The filling with water started by the Target volume, in which was pumped 1.7~m$^3$ of Gd doped water. Then, the filling of the Lateral veto tank was done, with 1.6~m$^3$ of pure water in the active volume and 3~m$^3$ in the surrounding external volume. Finally, the Top veto was filled with 1.4~m$^3$ of pure water. 

\subsection{DAQ Installation}

The installation of the readout electronics and the computing infrastructure in the project's container occurred during a few months, with dedicated trips to the nuclear site. In each visit, new elements were added to the DAQ system and a few necessary modifications on the software and firmware level were done. Consequently, after each trip, the operational conditions of the DAQ were slightly changed, mainly concerning adjustments of the Front-end electronics - which means the pulse offset and the threshold of discrimination - and the Trigger configuration.

In the first visits to the nuclear site, a reduced version of the DAQ was installed consisting of one local server, the Front-end electronics (5 modules $\times$ 8 channels), the NDAQ digitizers (5 modules $\times$ 8 channels), the Trigger and the HV Power Supply systems, as shown in figure \ref{fig:daq_overview}. During that period, some problems were noticed: one of the five digitizers was not accessible by the VME single board computer, limiting the system to 32 readout channels, and the Trigger logic was not taking into account the VETO signals to reject cosmic muons crossing the VETO detectors. The second problem had a major impact on the overall Trigger rate of the detector, since all cosmic muons crossing the detector volume were contributing to the rate. Because the NDAQ digitizers could not handle such high trigger rate, the experiment was still not ready to acquire data continuously and only short time runs were performed for debugging purposes.

\begin{figure}[ht!]
	\centering
	\includegraphics[scale=0.6]{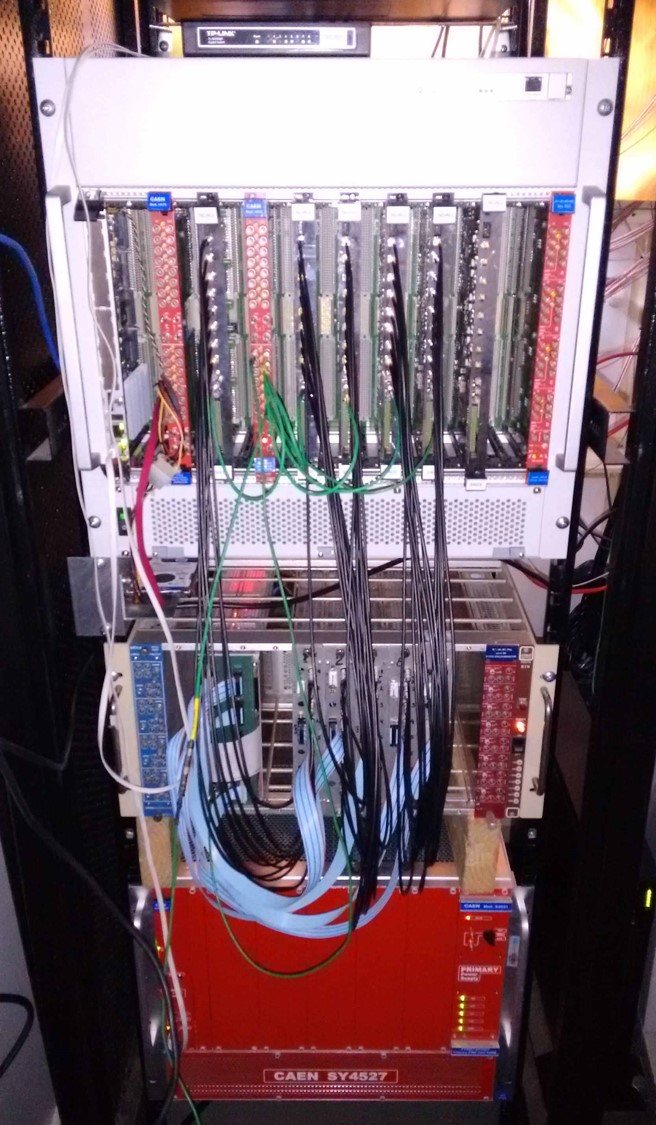}\\
	\caption{Data Acquisition electronics installed at the laboratory container located in the nuclear plant.}
	\label{fig:daq_pic}
\end{figure}

After a few more visits to the nuclear site laboratory, a first complete and operational version of the DAQ system was achieved as a result of three important actions: (i) new equipment was installed, (ii) a revised and corrected version of the Trigger firmware was configured and (iii) major fixes to the readout software were implemented. The new equipment included a high-power (3~kVA) UPS tower to provide stable AC power supply for all the DAQ elements, a commercial data storage unit with 16~TB capacity, and RaspberryPI-based electronics to allow remote configuration of the Front-end parameters - offset and threshold. After updating the Trigger firmware, the veto logic started to work properly. At this point, whenever two or more PMT hits occur at the same time window, any trigger pulse is blocked during the 2.5~$\mu$s veto window. Finally, with the VETO working as required, the overall detector trigger rate decreased to an average of 150~Hz (under the condition of at least four PMT hits in coincidence in the Target detector). A picture of the rack containing all the DAQ subsystems is shown in figure \ref{fig:daq_pic}. From top to bottom, one can see: (i) the VME crate containing the Single Board Computer, the NDAQ digitizers and the fanout modules, (ii) the NIM crate with the Front-end electronics and the Trigger module and (iii) the HV power supply subsystem.                

\section{Preliminary Measurements and Data Analysis}

Several experimental tests with the detector have been carried out during the commissioning campaigns in the nuclear facility, with the DAQ partially and fully installed. In the next paragraphs only the most relevant measurements, demonstrating the detector functionality and performance, will be described along with the achieved results.  

\begin{figure}[ht!]
	\centering
	\includegraphics[scale=0.32]{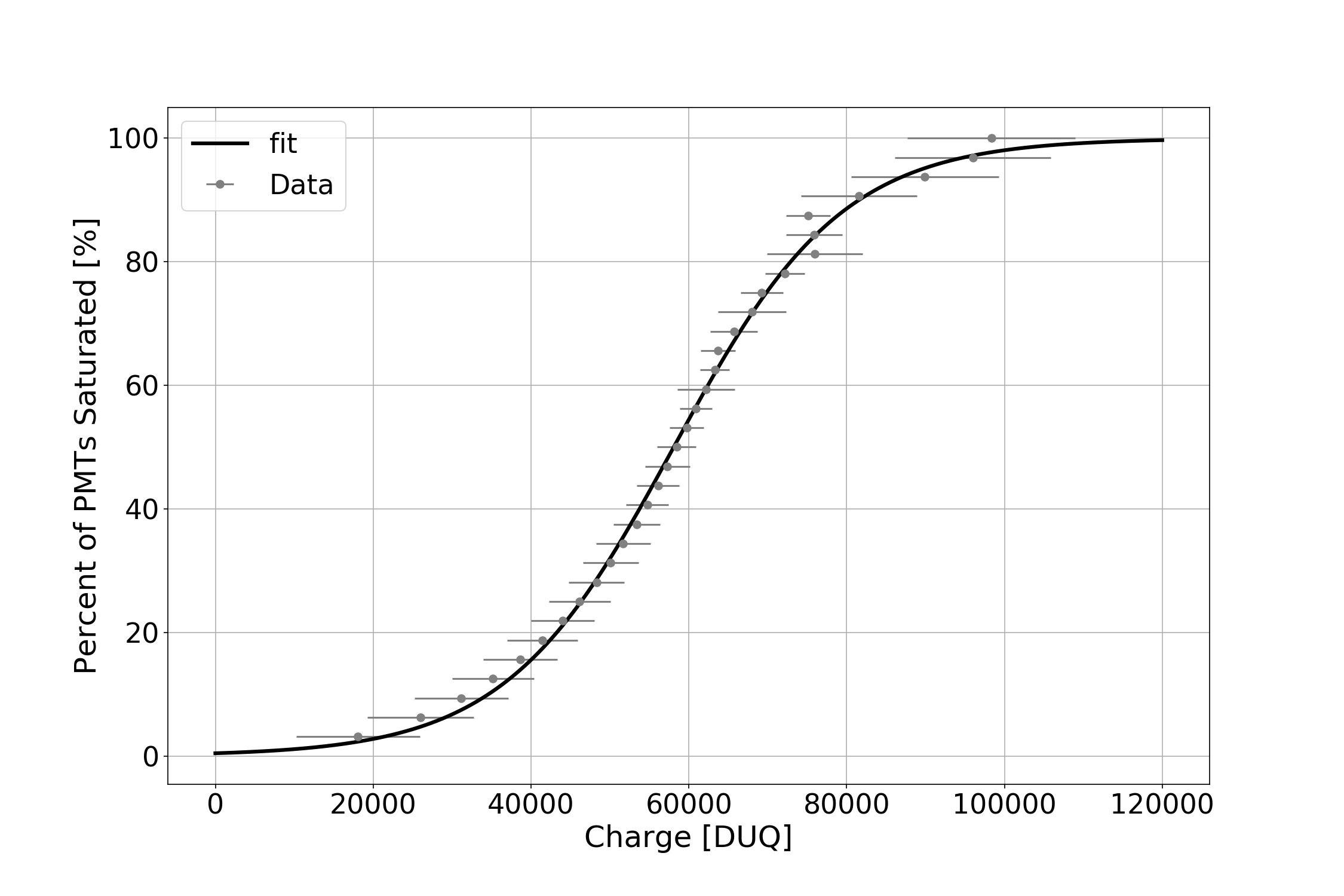}\\
	\caption{DAQ Electronics Saturation curve.}
	\label{fig:saturation}
\end{figure}

The first test of the electronics was validating the non-saturated charge region of the DAQ. This test is independent of the trigger system, depending only on the PMTs, High-Voltage system and Front-End gain. We defined as \textit{saturated} any pulse with at least 2 samples with the ADC saturated (using 8-bit configuration). The result can be seen in the figure \ref{fig:saturation} where each grey point represents the mean charge for which n-PMTs are saturated (shown in percent) and the solid line is a logistic curve fit. Although there is no precise calibration available during this commissioning phase, the inverse beta decay positrons and neutrons are expected to be detected with less than $3*10^4$ DUQ (digital unity of charge). The charge unity used during the commissioning phase, \textit{DUQ}, is calculated as the sum of the digital value of the ADCs samples.

\begin{figure}[ht!]
	\centering
	\includegraphics[scale=0.3]{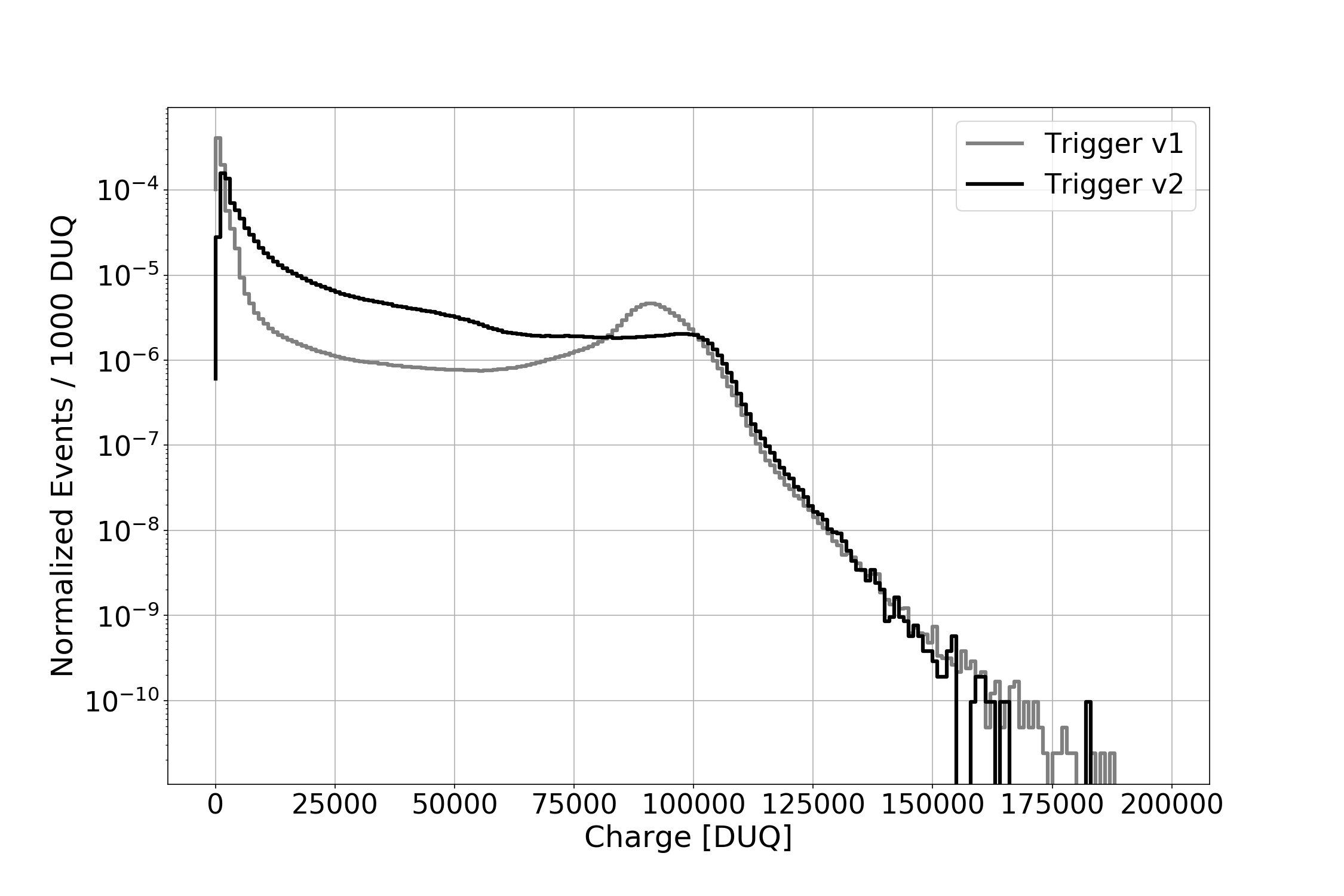}\\
	\caption{Charge Spectrum obtained with no cuts on the Target detector.}
	\label{fig:charge_hist}
\end{figure}

The first physics result obtained after several weeks of detector operation was the spectrum of particles (mostly muons) crossing and losing energy in the Target detector. Since the correction of the muon veto firmware was a drastic change in the way the detector was working, in figure \ref{fig:charge_hist} two spectra are shown, one for events before the firmware correction and the other after the correction. It is clear to notice that a bump between 75000 and 100000 DUQ units disappears for the events acquired with the veto working, as expected due to the removal of most of the vertical muons from the data. Those events are simply rejected by the veto system.      

\begin{figure}[ht!]
	\centering
	\includegraphics[scale=0.6]{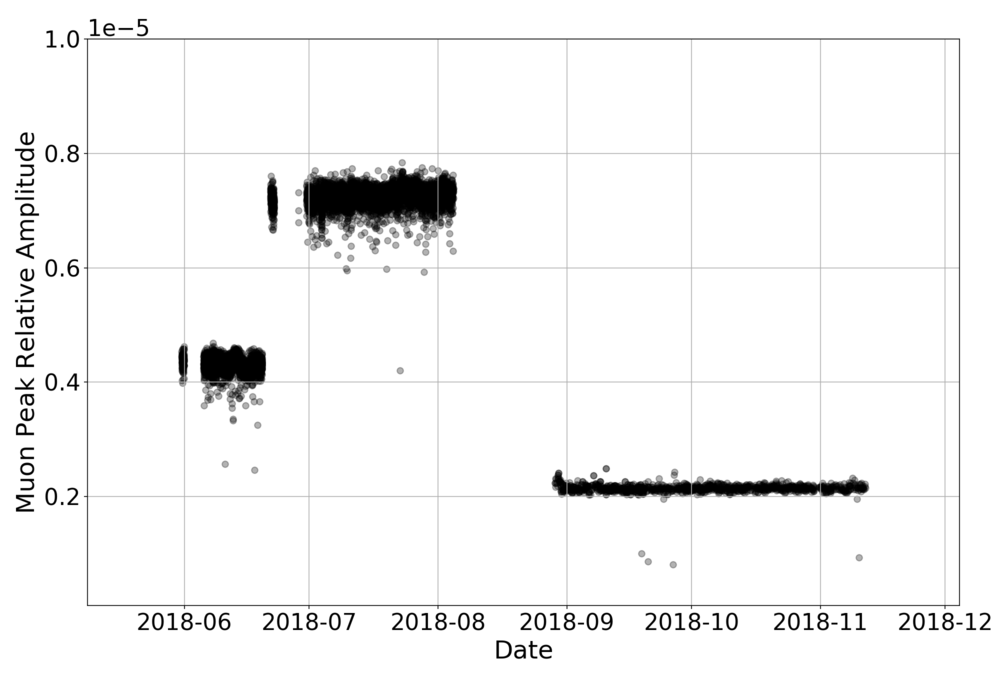}\\
	\caption{Relative amplitude of the muon peak (region showed between $7.5*10^4$ and $10^5$ on plot \ref{fig:charge_hist}).}
	\label{fig:muon_peak}
\end{figure}

A useful measurement to check and monitor the detector operation in the long run is the stability of the bump region amplitude: the high energy region dominated by crossing muons. Although this region is heavily saturated (more than 80\% according to figure \ref{fig:saturation}), the saturation effect shouldn't change over time, ensuring the use of this feature as a stability check. This result is presented in figure \ref{fig:muon_peak} for periods after each of the commissioning campaigns. In the last group of measurements, more recent weeks, the mean reconstructed amplitude is reduced, as expected, since the veto system is in action.

\begin{figure}[ht!]
	\centering
	\includegraphics[scale=0.6]{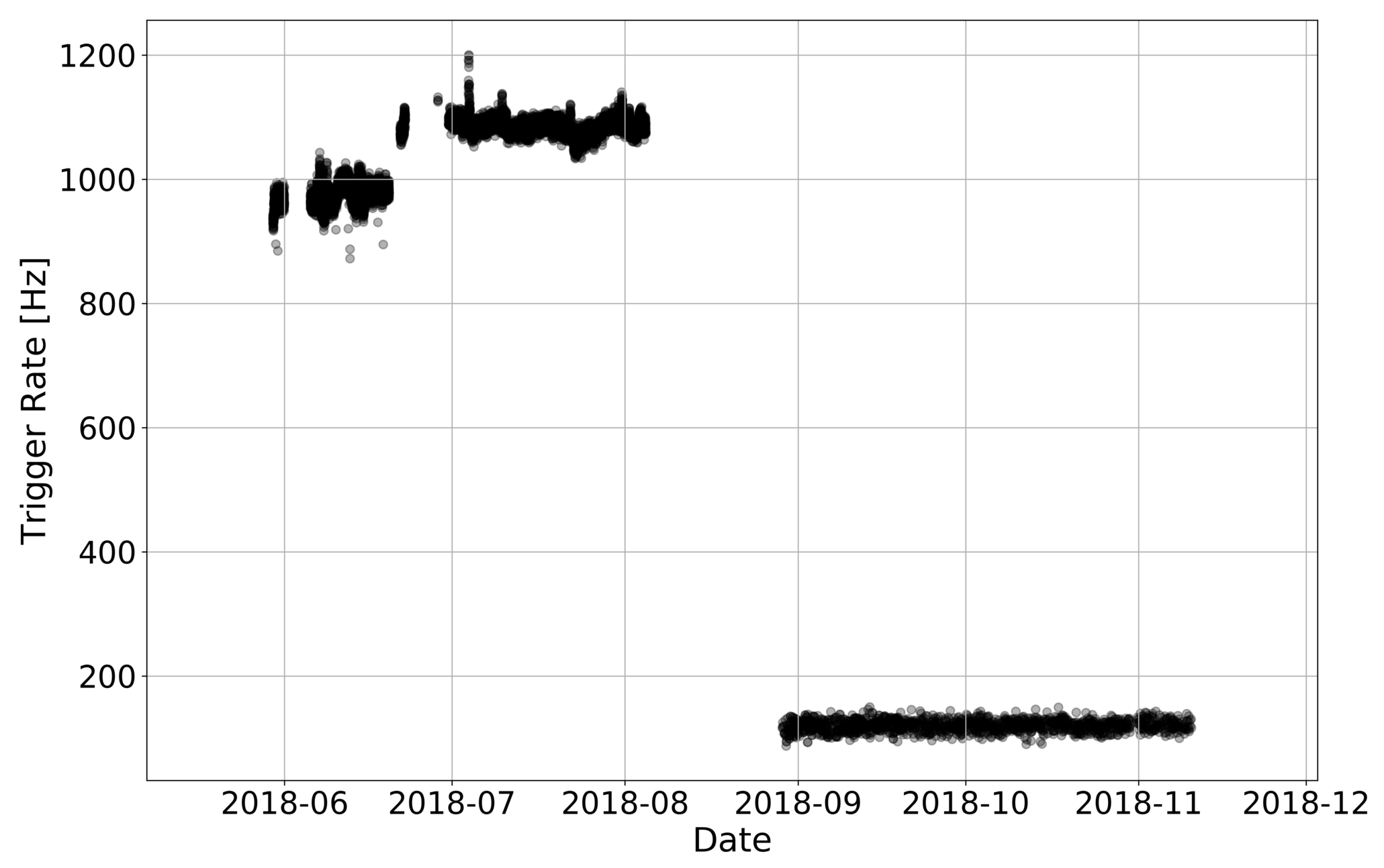}\\
	\caption{Trigger rate stability after each commissioning campaign.}
	\label{fig:trigger_rate}
\end{figure}

Trigger rate was also analyzed in this same context. In figure \ref{fig:trigger_rate} it is possible to see that the rate changes after each commissioning campaign, but between each of them the rate behavior remained stable. For the two first groups of measurements it should be emphasized that the Veto system was not working. That is the reason for the higher mean value, close to 1~kHz, when compared to the last group (starting in 2018-09). The small increase in the rate from the first to the second group may be explained by the adjustments done on the Front-end thresholds during the second DAQ installation campaign. Some channels on the Target detector were operating with threshold values higher than necessary. After the decrease of the threshold on those channels, it was expected to observe an increase in the individual trigger rate and, consequently, an increase in the overall trigger rate. After the last commissioning campaign, the overall trigger rate became stable around a mean value of 170~Hz. The rate of muons crossing the VETO detector is also monitored by the Readout Electronics and fluctuates around a mean value of 1~kHz.

\begin{figure}[ht!]
	\centering
	\includegraphics[scale=0.3]{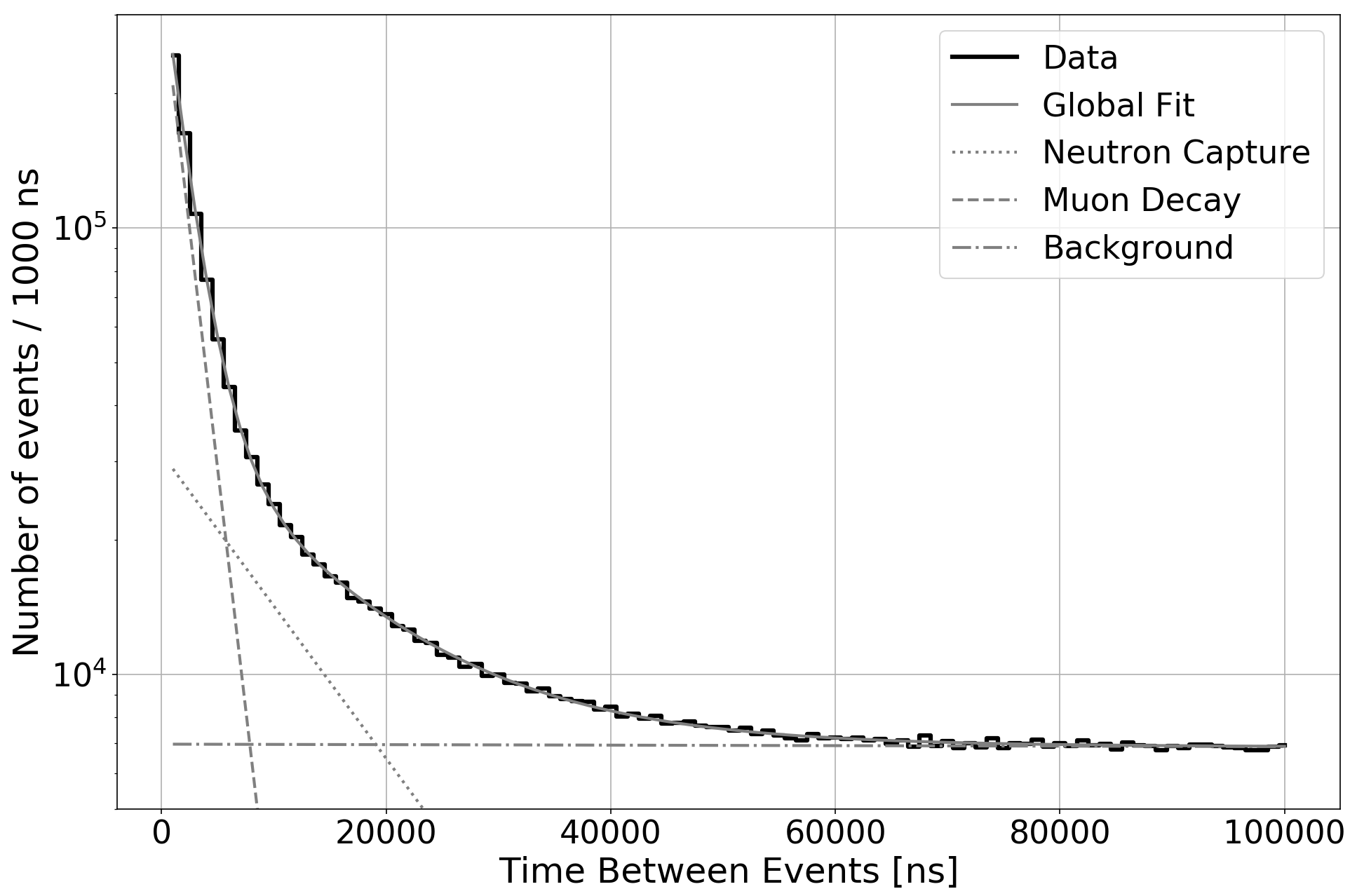}\\
	\caption{Time between events in the Target detector.}
	\label{fig:time_between}
\end{figure}

The distribution of time intervals between events in the Target detector is shown in figure \ref{fig:time_between}. In the range of tens of microseconds, there are three main different phenomena: neutron capture in the Gd, muon decay and random coincidences. In the plot of figure \ref{fig:time_between} two exponential curves were adjusted to fit the phenomena: the random coincidence was fixed from a fit with larger temporal scale. With these fitting results, it was possible to compute: (i) the event rate (from the background time constant), (ii) the mean time of Gd neutron capture (that is proportional to the amount of Gd in the Target detector water) and (iii) the mean lifetime of muons (an ensemble of positive and negative muons). The fitting result can be seen in table \ref{tab:time_constants}.\\

\begin{table}[ht!]
\centering
\begin{tabular}{c|c|c|}
\cline{2-3}
& \textbf{Time Constant (ns)} & \textbf{Fit Error (ns)} \\ \hline
\multicolumn{1}{|c|}{\textbf{Neutron Capture}} & 12700                       & 80                      \\ \hline
\multicolumn{1}{|c|}{\textbf{Muon decay}}      & 2015                        & 4                       \\ \hline
\multicolumn{1}{|c|}{\textbf{Backgroud}}       & $7.83*10^6$                     & $5*10^4$                   \\ \hline                
\end{tabular}
\caption{Fitting results from data acquired on last commissioning period.}
\label{tab:time_constants}
\end{table}

\begin{figure}[ht!]
  \subfloat[Charge spectrum of the two time coincidence windows]{%
    \includegraphics[width=7.5cm]{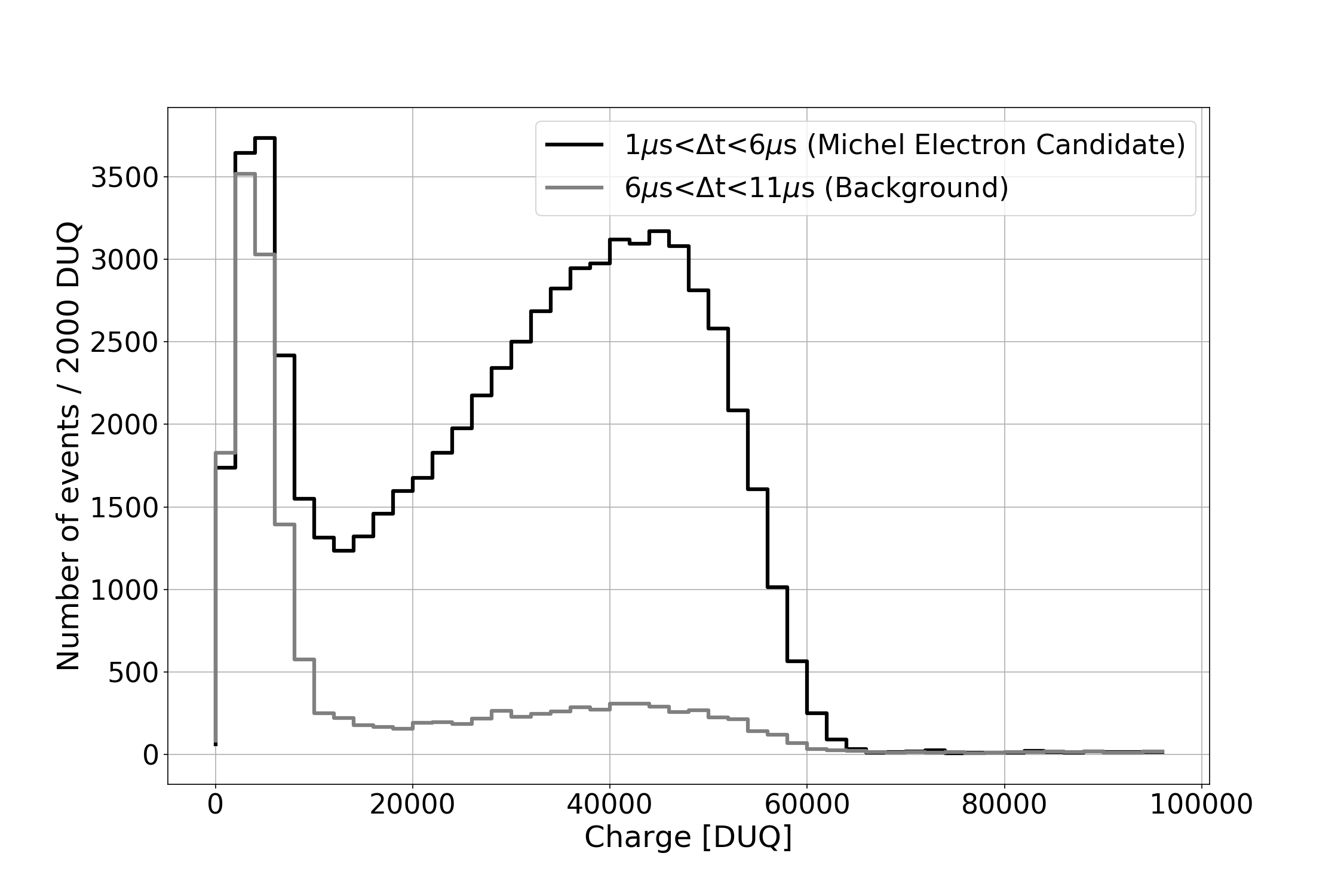} 
  } 
  \hfill 
  \subfloat[Background subtracted charge spectrum]{%
    \includegraphics[width=7.5cm]{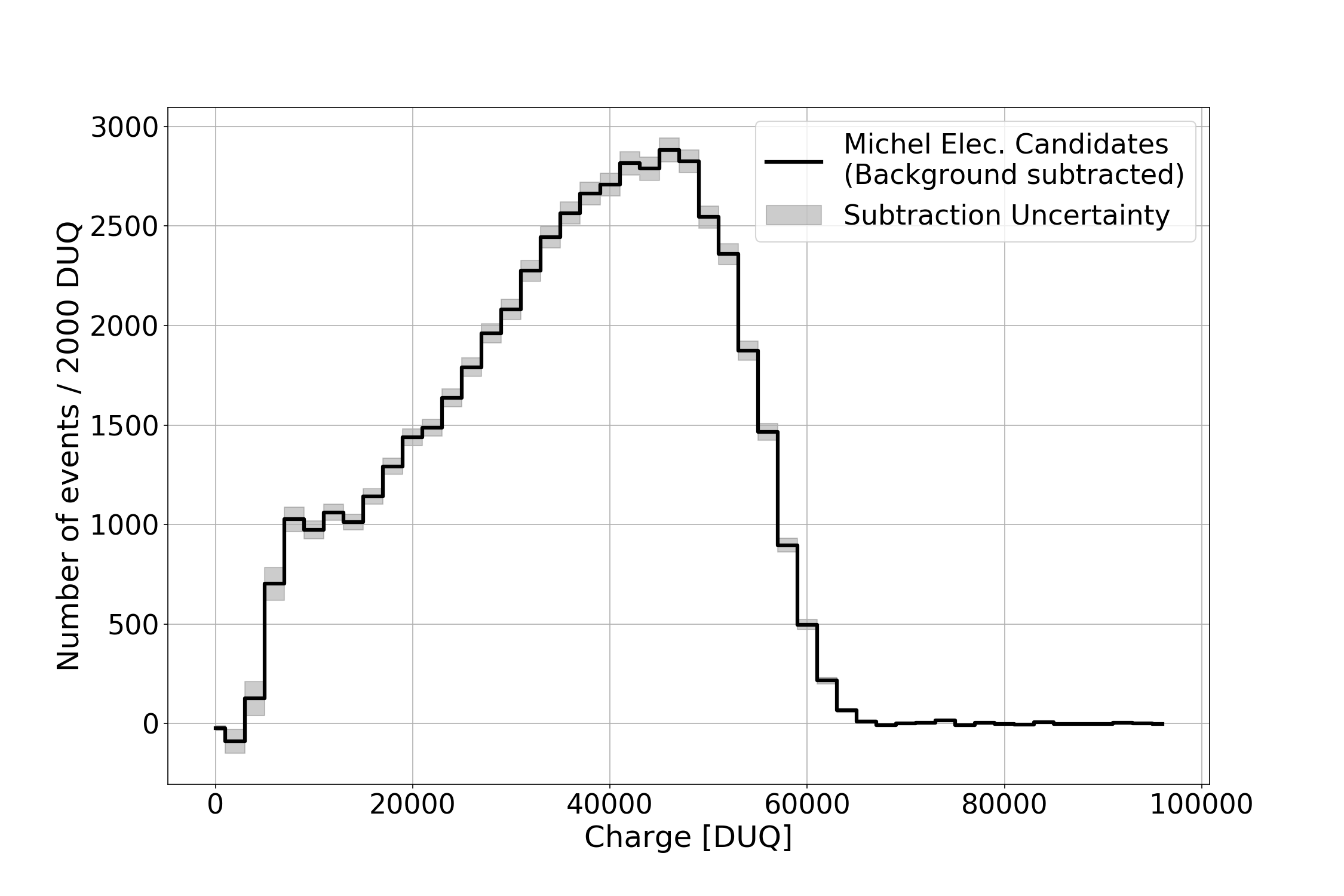} 
  } 
  \caption{Michel electrons candidates.} 
  \label{fig:michel}
\end{figure}

A selection of stopping muon candidates was made using events with charge between $5*10^4$ and $7.5*10^4$ DUQ and time coincidence between  $1~\mu s$ and  $6~\mu s$. A selection of background was made using a time coincidence between $6~\mu s$ and $11~\mu s$. After background exclusion, the charge of the second event of the coincidence can be plotted as the Michel Electron Candidates charge spectrum, as shown in figure \ref{fig:michel}. One should note that this plot has negative values in the low energy limit due to the background exclusion, which, as expected, is more dominated by background for this selection. 
The lack of sharpness on the end of the Michel Electron Candidates spectrum can be explained by the saturation of about $40\%$ of the events expected in that charge region, combined with the small size of the detector, which causes electrons to spill out of the target. In that energy region, the expected electron range in water is about 20~cm \cite{nist_estar}, making the detector center the only part able to work as an optimal calorimeter for Michel Electrons. We are working on a Data/Simulation comparison for the spectrum of Michel electrons, to explore its features as a calibration tool due the overlap with relevant energies from the events generated by the antineutrinos from the reactor.

\section{Conclusions}
The main aspects of the Neutrinos Angra detection system and its full commissioning are reported, with the description of the different tasks accomplished. Preliminary measurements are reported with the detector fully operational. First, the detector was fully assembled in the laboratory container located a few meters from the nuclear reactor protection. In a second period, after three campaigns that had lasted for a few months, the Data Acquisition system was fully installed, tested and put into operation. As the preliminary measurements presented confirm, the experiment is fully operational, the detector is working as expected and the DAQ is continuously taking data for further precision analysis towards the detection and identification of neutrinos.

\section*{Acknowledgment}
This work is supported by the Brazilian Ministry of Science, Technology, Innovation and Communications (MCTIC), FINEP, CNPq and FAPERJ. We also thank the staff of Eletrobr\'as Eletronuclear, the nuclear power plant operator. This study was financed in part by the Coordena\c{c}\~ao de Aperfei\c{c}oamento de Pessoal de N\'ivel Superior (CAPES) - Finance Code 001. This research used the computing resources and assistance of the John David Rogers Computing Center (CCJDR) in the Institute of Physics "Gleb Wataghin", University of Campinas. The Neutrinos Angra Collaboration would like to thank the invaluable contribution of Ademarlaudo Fran\c{c}a Barbosa, who passed away, whose work has been decisive for the results presented in this article. 


\end{document}